\documentclass[prl,twocolumn]{revtex4}
\usepackage{graphicx}
\begin{document}

\title{Axial charges of N(1535) and N(1650)
in lattice QCD with two flavors of dynamical quarks}

\author{Toru T. Takahashi and Teiji Kunihiro}

\affiliation{Yukawa Institute for Theoretical Physics, Kyoto University,
Kitashirakawa-Oiwakecho, Sakyo, Kyoto 606-8502, Japan}

\date{\today}

\begin{abstract}
We show the first lattice QCD results on the axial charge $g_A^{N^*N^*}$
of $N^*(1535)$ and $N^*(1650)$.
The measurements are performed with two flavors of
dynamical quarks employing the renormalization-group improved gauge
action at $\beta$=1.95 and the mean-field improved clover quark action
with the hopping parameters, $\kappa$=0.1375, 0.1390 and 0.1400. In
order to properly separate signals of $N^*(1535)$ and $N^*(1650)$,
we construct 2$\times$2 correlation matrices and diagonalize them.
Wraparound contributions in the correlator, which can be another source
of signal contaminations, are eliminated by imposing the Dirichlet
boundary condition in the temporal direction. We find that the axial
charge of $N^*(1535)$ takes small values as $g_A^{N^*N^*}\sim {\mathcal O}(0.1)$,
whereas that of $N^*(1650)$ is about 0.5,
which is found independent of quark masses
and consistent with the predictions by the naive nonrelativistic quark model.
\end{abstract}

\maketitle

{\bf Introduction.}
Chiral symmetry is an approximate global symmetry in QCD,
the fundamental theory of the strong interaction;
this symmetry together with its spontaneous breaking
has been one of the key ingredients 
in the low-energy hadron or nuclear physics.
Due to its spontaneous breaking, up and down quarks,
whose current masses are of the order of a few MeV,
acquire the large constituent masses of a few hundreds MeV,
and are consequently responsible for about 99\% of mass
of the nucleon and hence that of our world.
Thus one would say that chiral condensate $\langle \bar \psi \psi \rangle$,
the order parameter of the chiral phase transition,
plays an essential role in the 
hadron-mass genesis in the light quark sector.
On the other hand, chiral symmetry gets restored 
in systems where hard external 
energy scales such as high-momentum transfer,
temperature($T$), baryon density and so on exist,
owing to the asymptotic freedom of QCD.
Then, are all hadronic modes massless in such systems?
Can hadrons be massive even without non-vanishing chiral condensate?

An interesting possibility was suggested 
some years ago by DeTar and Kunihiro~\cite{DeTar:1988kn},
who showed that nucleons can be {\it massive even without
the help of chiral condensate}
due to the possible {\it chirally invariant mass terms},
which give {\it degenerated}
finite masses to the members in the chiral multiplet
(a nucleon and its parity partner)
even when chiral condensate is set to zero:
To show this for a finite-$T$ case, they introduced 
a linear sigma model which 
offers a nontrivial chiral structure in the baryon sector
and a mass-generation mechanism
completely and essentially different from that 
by the spontaneous chiral symmetry breaking.
Interestingly enough,
their chiral doublet model has recently become
a source of debate as a possible scenario of
{\it observed parity doubling in excited baryons
}~\cite{Jaffe:2005sq,Jaffe:2006jy,Glozman:2007jt,Jido:1999hd,Jido:2001nt,Lee:1972},
although their original work \cite{DeTar:1988kn} was 
supposed to be applied to finite-$T$ systems.

It is thus an intriguing problem 
to reveal the chiral structure
of excited baryons in the light quark sector
beyond model considerations.
One of the key observables which are sensitive to
the chiral structure of the baryon sector is axial charges~\cite{DeTar:1988kn}.
The axial charge of a nucleon $N$ is encoded in the three-point function
\begin{equation}
\langle N|
A_\mu^a
|N\rangle
=
\bar u
\frac{\tau^a}{2}
[
\gamma_\mu \gamma_5
g_A(q^2)
+
q_\mu \gamma_5
h_A(q^2)
]
u.
\end{equation}
Here, 
$A_\mu^a
\equiv
\bar Q \gamma_\mu \gamma_5 \frac{\tau^a}{2} Q$
is the isovector axial current.
The axial charge $g_A$ is defined by $g_A(q^2)$
with the vanishing transferred momentum $q^2=0$.
It is a celebrated fact
that the axial charge $g_A^{NN}$ of $N(940)$ is 1.26.
Though the axial charges in the chiral broken phase
can be freely adjusted with higher-dimensional possible terms
and cannot be the crucial clues for the chiral 
structure~\cite{Jaffe:2005sq,Jaffe:2006jy},
they would surely 
reflect the internal structure of baryons
and would play an important role in the clarification of the low-energy
hadron dynamics.

In this paper, 
we show the first unquenched lattice QCD study~\cite{Takahashi:2007ti}
of the axial charge $g_A^{N^*N^*}$ of $N^*(1535)$ and $N^*(1650)$.
We employ $16^3\times 32$ lattice with two flavors of dynamical quarks,
generated by CP-PACS collaboration~\cite{AliKhan:2001tx}
with the renormalization-group improved
gauge action and the mean-field improved clover quark action.
We choose the gauge configurations at $\beta=1.95$
with the clover coefficient $c_{\rm SW}=1.530$,
whose lattice spacing $a$ is determined as 0.1555(17) fm.
We perform measurements with 590, 680, and 680 gauge configurations
with three different hopping parameters for sea and valence quarks,
$\kappa_{\rm sea},\kappa_{\rm val}=0.1375,0.1390$ and $0.1400$,
which correspond to quark masses of $\sim$ 150, 100, 65 MeV and 
the related $\pi$-$\rho$ mass ratios are
$m_{\rm PS}/m_{\rm V}=0.804(1)$, $0.752(1)$ and $0.690(1)$, respectively.
Statistical errors are estimated by the jackknife method
with the bin size of 10 configurations.

Our main concern is the axial charges of the negative-parity 
nucleon resonances $N^*(1535)$ and $N^*(1650)$ in $\frac12^-$channel.
We then have to construct an optimal operator
which dominantly couples to $N^*(1535)$ or $N^*(1650)$.
We employ the following two independent nucleon fields,
$
N_1(x)\equiv \varepsilon_{\rm abc}u^a(x)(u^b(x)C\gamma_5 d^c(x))
$
and
$
N_2(x)\equiv \varepsilon_{\rm abc}\gamma_5 u^a(x)(u^b(x)C d^c(x)),
$
in order to construct correlation matrices
and to separate signals of $N^*(1535)$ and $N^*(1650)$.
(Here, $u(x)$ and $d(x)$ are Dirac spinor for u- and d- quark,
respectively, and $a,b,c$ denote the color indices.)
Even after the successful signal separations,
there still remain several signal contaminations
mainly because lattices employed 
in actual calculations are finite systems:
Signal contaminations 
 ({\it a}) {\it by scattering states},
 ({\it b}) {\it by wraparound effects}.

{Comment to  ({\it a}) :}
Since our gauge configurations are unquenched ones,
the negative parity nucleon states could decay to $\pi$ and N,
and their scattering states could come into the spectrum.
The sum of the pion mass $M_\pi$ and the nucleon mass $M_N$
is however in our setups heavier than the masses of the lowest two states
(would-be $N^*(1535)$ and $N^*(1650)$)
in the negative parity channel.
We then do not suffer from any scattering-state signals.

{Comment to  ({\it b}) :}
The other possible contamination is wraparound 
effects~\cite{Takahashi:2005uk}.
Let us consider a two-point baryonic correlator 
$\langle N^*(t_{\rm snk}) \bar N^*(t_{\rm src})\rangle$
in a Euclidean space-time.
Here, the operators $N^*(t)$ and $\bar N^*(t)$ 
have nonzero matrix elements,
$\langle 0|N^*(t)|N^*\rangle$ and
$\langle N^*|\bar N^*(t)|0\rangle$,
and couple to the state $|N^*\rangle$.
Since we perform unquenched calculations,
the excited nucleon $N^*$ can decay into $N$ and $\pi$,
and even when we have no scattering state $|N+\pi\rangle$,
we could have another ``scattering states''.
The correlator
$\langle N^*(t_{\rm snk}) \bar N^*(t_{\rm src})\rangle$
can still accommodate, for example, the following term.
\begin{eqnarray}
&&\langle \pi|N^*(t_{\rm snk})|N\rangle
\langle N| \bar N^*(t_{\rm src})|\pi\rangle \nonumber \\
&\times&e^{-E_N(t_{\rm snk}-t_{\rm src})}\times
e^{-E_\pi (N_t-t_{\rm snk}+t_{\rm src})}.
\end{eqnarray}
Here, $N_t$ denotes the temporal extent of a lattice.
Such a term is quite problematic and mimic a fake plateau
at $E_N-E_\pi$ in the effective mass plot
because it behaves as $\sim e^{-(E_N-E_\pi)(t_{\rm snk}-t_{\rm src})}$.
Although these contaminations disappear when one employ enough large-$N_t$
lattice, our lattices do not have so large $N_t$.
In order to eliminate such contributions,
we impose the Dirichlet condition on the temporal
boundary for valence quarks,
which prevents valence quarks from going over the boundary.
Though the boundary is still transparent for the states
with the same quantum numbers as vacuum, {\it e.g.} glueballs,
such contributions will be suppressed
by the factor of $e^{-E_GN_t}$ and we neglect them in this paper.
(Wraparound effects can be found even in quenched 
calculations~\cite{Takahashi:2005uk}.)

{\bf Formulation.}
We here give a brief introduction to our 
formulation~\cite{Takahashi:2005uk,Burch:2006cc}.
Let us assume that  we have a set of $N$ independent operators,
$O_{\rm snk}^I$ for sinks and $O_{\rm src}^{I\dagger}$ for sources.
We can then construct an $N\times N$ correlation matrix
${\cal C}^{IJ}(T)
\equiv
\langle O_{\rm snk}^I(T)O_{\rm src}^{J\dagger}(0)\rangle
=C_{\rm snk}^\dagger\Lambda(T)C_{\rm src}
$.
Here, 
$
(C^\dagger_{\rm snk})_{Ii}\equiv \langle 0 | O_{\rm snk}^I | i \rangle
$
and
$
(C_{\rm src})_{jI}\equiv \langle j | O_{\rm src}^{J \dagger} | 0 \rangle
$
are general matrices,
and $\Lambda(T)_{ij}$ is a diagonal matrix given by
$
\Lambda(T)_{ij}\equiv \delta_{ij} e^{- E_iT}.
$
The optimal source and sink operators,
${\cal O}_{\rm src}^{i \dagger}$ and 
${\cal O}_{\rm snk}^{i}$,
which couple dominantly (solely in the ideal case) to $i$-th lowest state,
are obtained as
$
{\cal O}_{\rm src}^{i \dagger}=\sum_J
O^{J\dagger}_{\rm src} (C_{\rm src})^{-1}_{Ji}
$
and
$
{\cal O}_{\rm snk}^{i}=\sum_J
(C^\dagger_{\rm snk})^{-1}_{iJ} O^J_{\rm snk},
$
since $(C^\dagger_{\rm snk})^{-1} {\cal C}(T)(C_{\rm src})^{-1}
=\Lambda(T)$
is diagonal.
Besides overall constants,
$(C_{\rm src})^{-1}$ and $(C^\dagger_{\rm snk})^{-1}$
are obtained as the right and left eigenvectors
of 
${\cal C}^{-1}(T+1){\cal C}(T)$ and ${\cal C}(T){\cal C}(T+1)^{-1}$,
respectively.

The zero-momentum-projected point-type operators,
\begin{equation}
N_1(t)\equiv \sum_{\bf x}\varepsilon_{\rm abc}
u^a({\bf x},t)(u^b({\bf x},t)C\gamma_5 d^c({\bf x},t))
\end{equation}
and
\begin{equation}
N_2(t)\equiv \sum_{\bf x}\varepsilon_{\rm abc}\gamma_5 
u^a({\bf x},t)(u^b({\bf x},t)C d^c({\bf x},t)),
\end{equation}
are chosen for the sinks.
For the sources, we employ the following wall-type operators in the
Coulomb gauge,
\begin{equation}
\overline{N_1}(t)\equiv
\sum_{{\bf x_1},{\bf x_2},{\bf x_3}}
\varepsilon_{\rm abc}\bar u^a({\bf x_1},t)
(\bar u^b({\bf x_2},t)C\gamma_5 \bar d^c({\bf x_3},t))
\end{equation}
and
\begin{equation}
\overline{N_2}(t)\equiv
\sum_{{\bf x_1},{\bf x_2},{\bf x_3}}
\varepsilon_{\rm abc}\gamma_5 \bar u^a({\bf x_1},t)
(\bar u^b({\bf x_2},t)C \bar d^c({\bf x_3},t)).
\end{equation}
The parity is flipped by multiplying the operator by $\gamma_5$;
$N^+_i(t)\equiv          N_i(t)$ and
$N^-_i(t)\equiv \gamma_5 N_i(t)$.
The optimized sink (source) operators
${\cal N}^\pm_i$ ($\overline{{\cal N}^\pm_i}$),
which couple dominantly to the $i$-th lowest state are constructed as
\begin{eqnarray}
{\cal N}_i^\pm(t)&=&
N^\pm_1(t)
+\left[
{(C^{\pm \dagger}_{\rm snk})^{-1}_{i2}} /
{(C^{\pm \dagger}_{\rm snk})^{-1}_{i1}}
\right]
N^\pm_2(t) \\
&\equiv&
N^\pm_1(t)
+L_i^\pm
N^\pm_2(t),
\end{eqnarray}
and
\begin{eqnarray}
\overline{{\cal N}^\pm_i}(t)&=&
\overline{N_1^\pm}(t)
+\left[
{(C^\pm_{\rm src})^{-1}_{2i}} /
{(C^\pm_{\rm src})^{-1}_{1i}}
\right]
\overline{N_2^\pm}(t) \\
&\equiv&
\overline{N_1^\pm}(t)
+R_i^\pm
\overline{N_2^\pm}(t).
\end{eqnarray}

\begin{figure}[h]
\begin{center}
\includegraphics[scale=0.3]{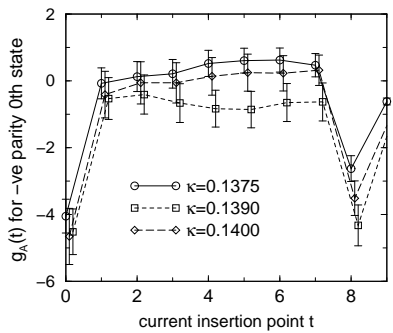}
\end{center}
\caption{\label{3pointfunc}
The non-renormalized axial charge of $N^*(1535)$,
$g_{A}^{-0{\rm [lat]}}(t)$,
as a function of the current insertion time $t$.
}
\end{figure}

Now that we have constructed optimized operators,
we can easily compute the (non-renormalized) vector and axial charges
$g_{V,A}^{\pm{\rm [lat]}}$ for the positive- and negative-parity nucleons
via three-point functions with the so-called sequential-source 
method~\cite{Sasaki:2003jh}.
In practice, we evaluate $g_{V,A}^{\pm{\rm [lat]}}(t)$ defined as
\begin{equation}
g_{V,A}^{\pm{\rm [lat]}}(t)
=
\frac
{
{\rm Tr}\ \Gamma_{V,A}
\langle B(t_{\rm snk})
J_\mu^{V,A}(t)
\overline B(t_{\rm src}) \rangle
}
{
{\rm Tr}\ \Gamma_{V,A}
\langle B(t_{\rm snk})
\overline B(t_{\rm src}) \rangle
},
\end{equation}
and extract $g_{V,A}^{\pm{\rm [lat]}}$ by the fit
$g_{V,A}^{\pm{\rm [lat]}}=g_{V,A}^{\pm{\rm [lat]}}(t)$ in the plateau region.
$B(t)$ denotes the (optimized) interpolating field for nucleons,
and $\Gamma_{V,A}$ are $\gamma_\mu \frac{1+\gamma_4}{2}$
and $\gamma_\mu \gamma_5 \frac{1+\gamma_4}{2}$, respectively.
$J_\mu^{V,A}(t)$ are the vector and the axial vector currents 
inserted at $t$.
We show in Fig.~\ref{3pointfunc}
$g_{A}^{-0{\rm [lat]}}(t)$ for $N^*(1535)$
as a function of the current insertion time $t$.
They are rather stable around $t_{\rm src}<t<t_{\rm snk}$.

We finally reach the renormalized charges
$g_{A,V}^\pm=\widetilde Z_{A,V}g^{\pm{\rm [lat]}}_{A,V}$
with the prefactors
$\widetilde Z_{A,V}
\equiv
2\kappa
u_0
Z_{V,A}
\left(
1+b_{V,A}\frac{m}{u_0}
\right)
$,
which are 
estimated with the values listed in Ref.~\cite{AliKhan:2001tx}.

\begin{table*}[ht]
\caption{
\label{fittedval}
{\bf upper table:}
The fitted values of $L_{1,2}^\pm$ and $R_{1,2}^\pm$
for the ground and the 1st excited states in positive- and negative-parity
channels are listed.
$M_\pi$ and $E_{1,2}^\pm$ 
denote the pion mass and the ground- and the 1st excited-state
energies for the positive- and the negative-parity channels
for each $\kappa$ in the lattice unit.
($a=0.1555$fm and $a^{-1}=1.267$GeV.)
The row ``{\rm C.L.}'' shows the values at the chiral limit.
{\bf lower table:}
The non-renormalized vector and axial charges
for $n$-th positive- and negative-parity nucleons
$g_{V,A}^{n\pm (u,d){\rm [lat]}}$
are listed. The superscripts $(u)$ and $(d)$
denote the u- and d-quark contributions, respectively.
The total axial and vector charges,
$g_V^{n\pm {\rm [lat]}}\equiv 
g_V^{n\pm (u){\rm [lat]}} - g_V^{n\pm (d){\rm [lat]}}$
and
$g_A^{n\pm {\rm [lat]}}\equiv 
g_A^{n\pm (u){\rm [lat]}} - g_A^{n\pm (d){\rm [lat]}}$,
as well as the renormalization factor and the improvement coefficients
$\widetilde Z_{V,A}
\equiv
2\kappa
u_0
Z_{V,A}
\left(
1+b_{V,A}\frac{m}{u_0}
\right)
$~\cite{AliKhan:2001tx}
for vector and axial currents are also listed.
$g_{V,A}^{n\pm {\rm}}\equiv\widetilde Z_{V,A} g_{V,A}^{n\pm {\rm [lat]}}$
denote the renormalized charges.
}
\newcommand{\m}{\hphantom{$-$}}
\begin{center}
\begin{tabular*}{1.0\textwidth}
{@{\extracolsep{\fill}}cccccccccccccc}
\hline
\hline
$\kappa$ 
& $L_1^+$ & $R_1^+$ & $L_1^-$ & $R_1^-$
& $L_2^+$ & $R_2^+$ & $L_2^-$ & $R_2^-$ 
& $M_\pi$ 
& $E_1^+$ & $E_1^-$ & $E_2^+$ & $E_2^-$ \\ \hline
0.1375 
& $-$0.4341  & $-$0.4573   & 0.0355      & \m0.0126
& $-$1353    & $-$314.1    & $-$1.432    & $-$1.302
& 0.8985(5) 
& 1.696(1)   & 2.137(10)   & 2.524(53)   & 2.141(14) \\
0.1390 
& $-$0.4526  & $-$0.4552   &  0.1115     & $-$0.2036
& $-$845.9   & $-$228.1    & $-$2.729    & $-$1.084
& 0.7351(5) 
& 1.459(1)   & 1.854(13)   & 2.162(44)   & 1.908(17) \\
0.1400 
& $-$0.1605  & $-$0.3552   & 0.0990      & $-$0.0151
& $-$408.9   & $-$143.6    & $-$1.510    & $-$1.038
& 0.6024(6) 
& 1.270(2)   & 1.665(15)   & 2.046(67)   & 1.733(25) \\ \hline
C.L.
& -        & -         & -         & -
& -        & -         & -         & -
& - 
& 0.936(3) & 1.277(25) & 1.570(109)& 1.411(38) \\ \hline \hline
\end{tabular*}

\vspace{.2cm}

\begin{tabular*}{1.0\textwidth}
{@{\extracolsep{\fill}}ccccccccccc}
\hline
\hline
$\kappa$ & 
${g^{0+{\rm [lat]}}_V}$ &
${g^{0-{\rm [lat]}}_V}$ & 
${g^{0+(u){\rm [lat]}}_A}$ & ${g^{0+(d){\rm [lat]}}_A}$ & ${g^{0+{\rm [lat]}}_A}$ &
${g^{0+{\rm}}_V}$ & ${g^{0-{\rm}}_V}$ & ${g^{0+{\rm}}_A}$ &
$\widetilde Z_V$ & $\widetilde Z_A$
 \\ \hline
0.1375 &
4.208( 8)   &
3.844( 76)  & 
3.852( 42) & $-$1.073(49)  &   4.925( 24) &
1.045(1) &    0.989( 19) &   1.247(8) &
0.2530 & 0.2576 \\
0.1390 &
4.492(10)  &
4.152(160) & 
3.978( 94) & $-$1.244(44) &    5.222(126) &
1.089(1) &    1.036( 60) &   1.261(7) &
0.2446 & 0.2491 \\
0.1400 &
4.663( 9)  &
4.380(206) & 
3.952(136) & $-$1.150(55)  &   5.102(145) &
1.115(2) &    1.048(111) &   1.261(8) &
0.2390 & 0.2434 \\ \hline \hline
\end{tabular*}

\vspace{.2cm}

\begin{tabular*}{1.0\textwidth}
{@{\extracolsep{\fill}}ccccccccc}
\hline
\hline
$\kappa$ & 

${g^{0-(u){\rm [lat]}}_A}$ & ${g^{0-(d){\rm [lat]}}_A}$ & ${g^{0-{\rm [lat]}}_A}$ &
${g^{1-(u){\rm [lat]}}_A}$ & ${g^{1-(d){\rm [lat]}}_A}$ & ${g^{1-{\rm [lat]}}_A}$ &
${g^{0-{\rm}}_A}$ &
${g^{1-{\rm}}_A}$ 
\\ \hline
0.1375 &
\m0.336(194) & $-$0.257(118) & \m0.592(226) &
  3.308(234) &    1.189(209) &   2.119(359) &
\m0.152(58)  &
0.546(093) \\
0.1390 &
$-$0.710(251) & \m0.081(119) & $-$0.791(272) &
   3.423(495) &   1.243(420) &    2.180(730) &
$-$0.197(68)  &
0.543(182) \\
0.1400 &
\m0.189(257) & $-$0.129(178) & \m0.318(297) &
  3.530(516) &    1.339(405) &   2.190(676) &
\m0.077(72)  &
0.533(165) \\
\hline \hline
\end{tabular*}

\end{center}
\end{table*}

\begin{figure}[h]
\begin{center}
\includegraphics[scale=0.3]{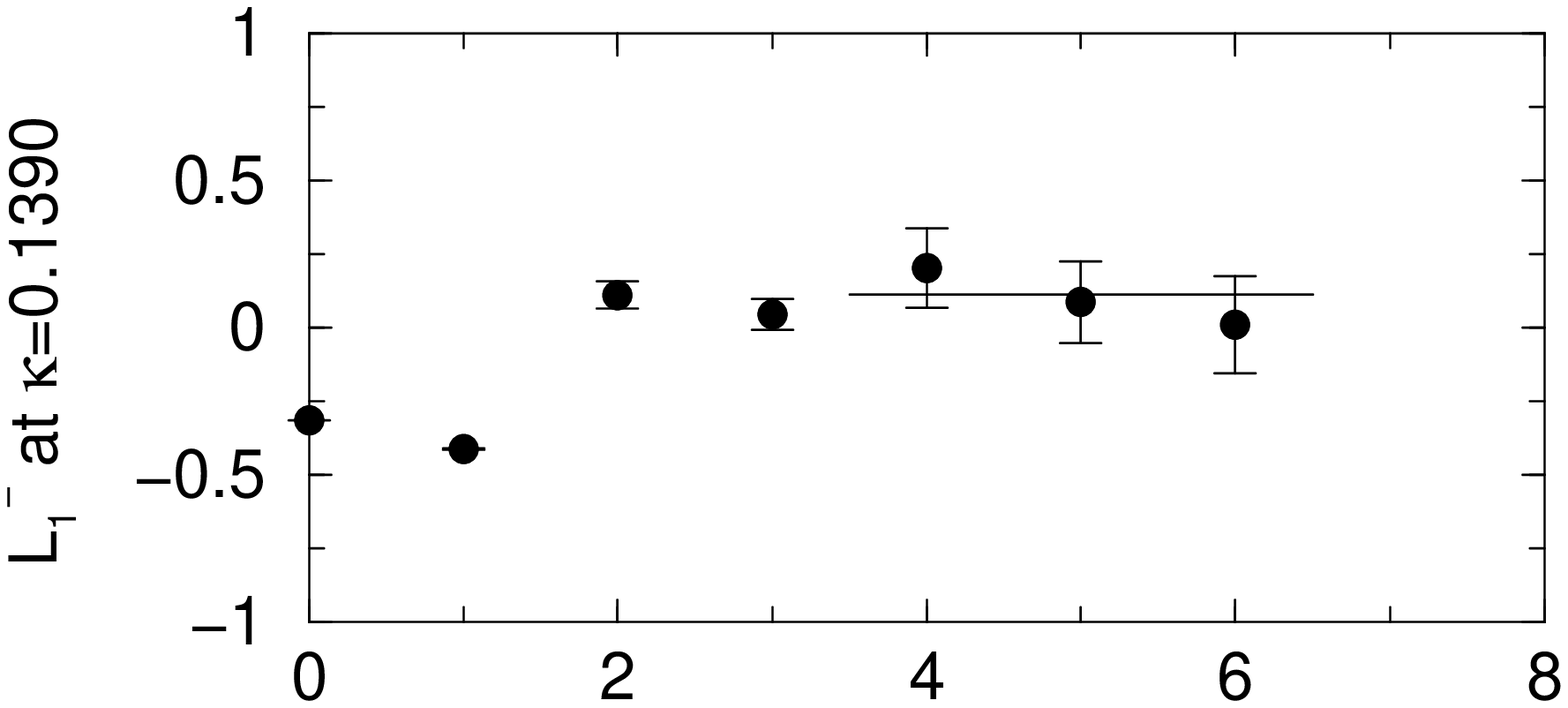}
\includegraphics[scale=0.3]{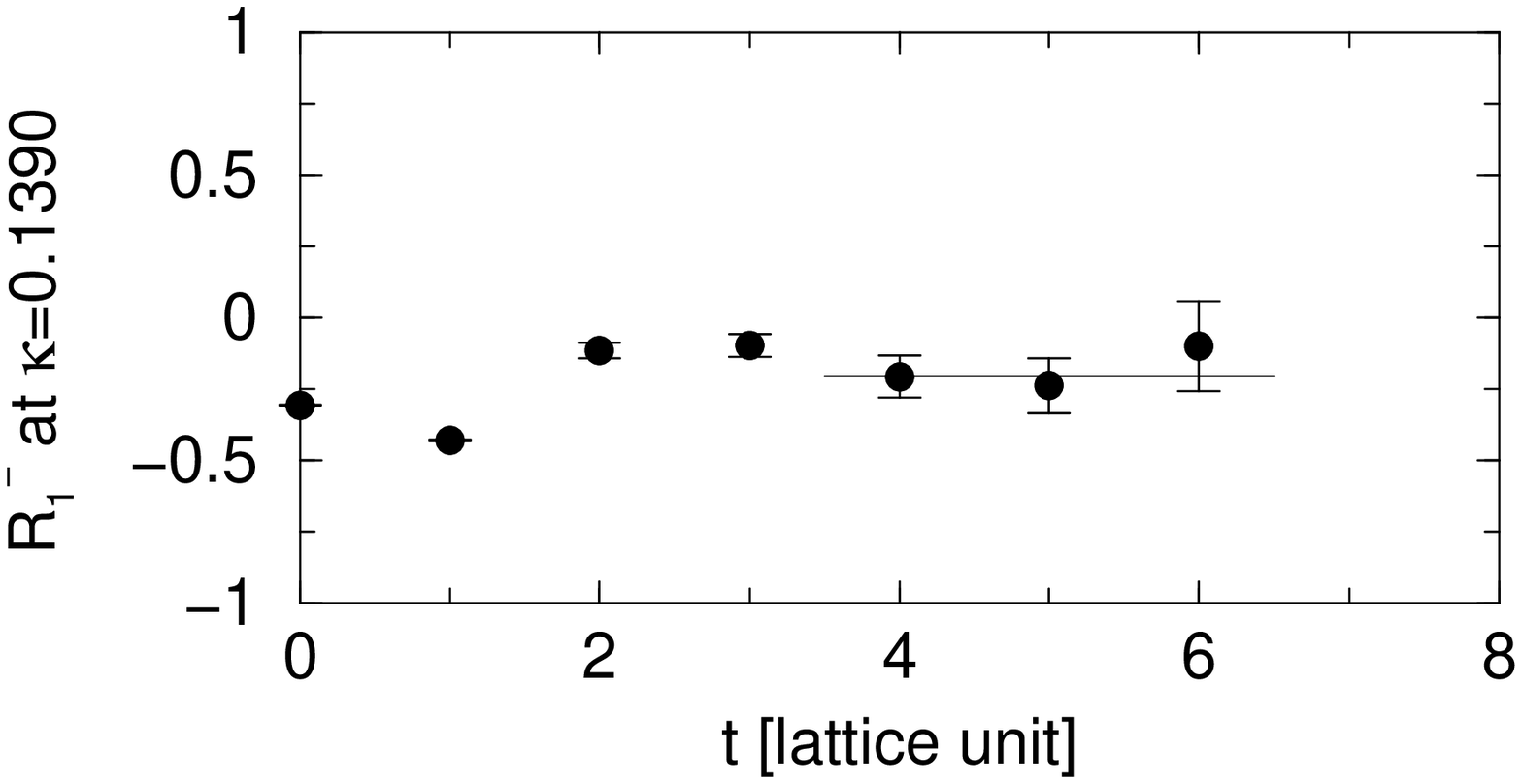}
\end{center}
\caption{\label{optparam}
As typical examples,
$L_1^-$ and $R_1^-$ obtained at $\kappa$=0.1390 are plotted.
}
\end{figure}

We show the fitted values of $L_{1,2}^\pm$ and $R_{1,2}^\pm$
in Table~\ref{fittedval}.
$L(T)$ and $R(T)$ are rather stable and show a plateau 
from relatively small value of $T$ ($T\sim 2$),
which is the same tendency as that found in Ref.~\cite{Burch:2006cc}.
We plot in Fig.~\ref{optparam} $L_i^\pm(T)$ and $R_i^\pm(T)$ obtained
at $\kappa$=0.1390, for the purpose of reference.
The energies $E_{1,2}^\pm$ are extracted 
from two-point correlation functions by the exponential fit as
$\langle {\cal N}^\pm_i(t_{\rm src}+T)
\overline{{\cal N}^\pm_i}(t_{\rm src}) \rangle
= C \exp(-E_i^\pm T)$ in the large-$T$ region.
The value at each hopping parameter is found to coincide with
that in the original paper by the CP-PACS 
collaboration~\cite{AliKhan:2001tx}, 
with deviations of 0.1\% to 1\%.
We here perform simple linear chiral extrapolations for $E_{1,2}^\pm$.
The chirally extrapolated values
as well as those at each hopping parameter
for $E_{1,2}^\pm$ in lattice unit
are listed in Table~\ref{fittedval}.
Although the mass $E_1^+$ of the ground-state positive-parity nucleon
at the chiral limit is overestimated in our analysis
($a^{-1}=1.267$ GeV),
this failure comes from our simple linear fit.

\begin{figure}[h]
\begin{center}
\includegraphics[scale=0.3]{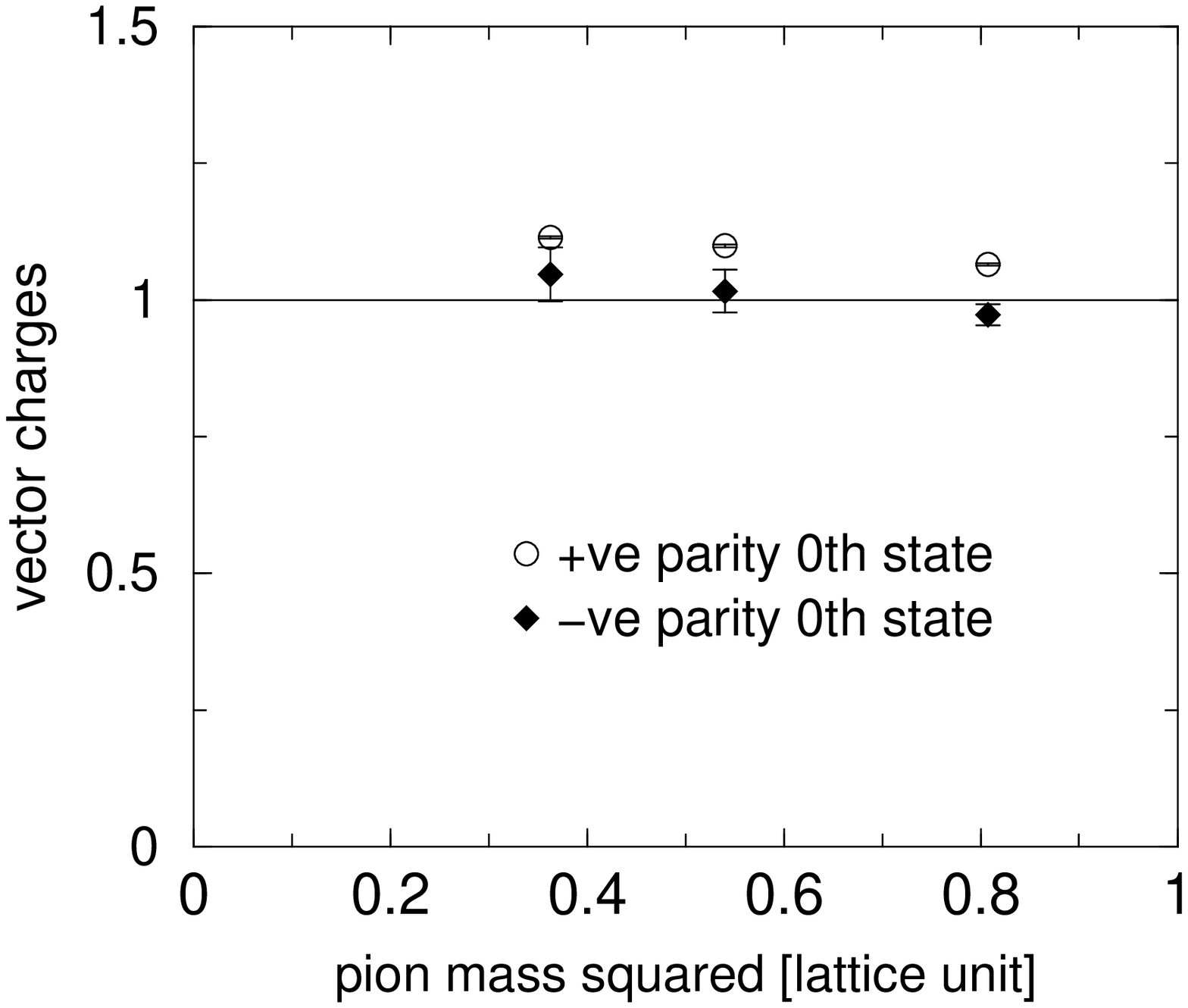}
\includegraphics[scale=0.3]{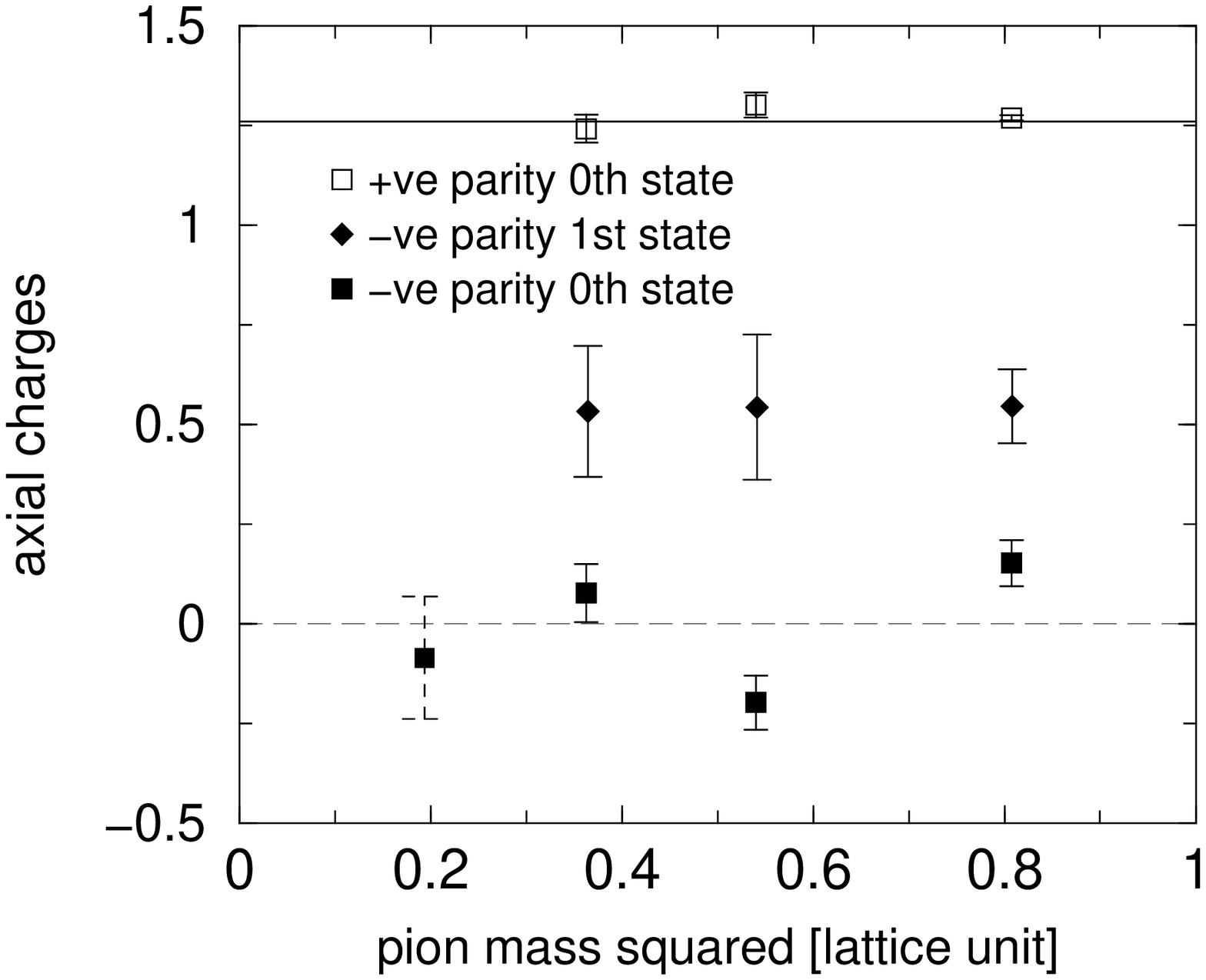}
\end{center}
\caption{\label{AxialVectorC}
The renormalized vector and axial charges of the positive- and the 
negative-parity nucleons are plotted
as a function of the squared pion mass $m_\pi^2$.
{\bf upper panel}:
The results of the vector charges.
The solid line is drawn at ${g}_V=1$ for reference.
{\bf lower panel}:
The results of the axial charges.
The solid line is drawn at ${g}_A=1.26$ and 
the dashed line is drawn at ${g}_A=0$.
}
\end{figure}

{\bf Results.}
We first take a stock of
the vector charges $g_V^{0\pm}$ of the ground-state 
positive- and negative-parity
nucleons as well as the axial charge $g_A^{0+}$ of the 
ground-state positive-parity nucleon,
which are well known and can be the references.
We show $g_V^{0\pm}$,
the vector charges of the positive- and the negative-parity nucleons
obtained with three hopping parameters $\kappa$=0.1375, 0.1390 and 0.1400,
in the upper panel in Fig.~\ref{AxialVectorC},
where the vertical axis denotes $g_V^{0\pm}$
and the horizontal one the squared pion masses.
(These values are also listed in Table~\ref{fittedval}.)
The vector charges should be unity
if the charge conservation is exact,
whereas we can actually find about 10\% deviations
in Table~\ref{fittedval} or in the upper panel in Fig.~\ref{AxialVectorC}.
Such unwanted deviations are considered to arise
due to the discretization errors:
The present lattice spacing is about 0.15 fm,
which is far from the continuum limit.
In fact, the decay constants obtained with the same setup as ours
deviate from the continuum values by ${\cal O}(10)$\%.
We should then count at least 10\% ambiguities in our results.
The axial charge $g_A^{0+}$
of the positive parity nucleon is also shown in the lower panel 
in Fig.~\ref{AxialVectorC}.
As found in the previous lattice studies,
the axial charge of the positive parity nucleon
shows little quark-mass dependence,
and they lie around the experimental value 1.26.

We finally show the axial charges of the negative-parity nucleon
resonances in the lower panel in Fig.~\ref{AxialVectorC}.
One finds at a glance that 
the axial charge $g_A^{0-}$
of $N^*(1535)$ takes quite small value,
as $g_A^{0-}\sim {\mathcal O}(0.1)$
and that even the sign is quark-mass dependent.
While the wavy behavior might come from
the sensitiveness of $g_A^{0-}$ to quark masses,
this behavior may indicate that
$g_A^{0-}$ is rather consistent with zero.
These small values are not the consequence
of the cancellation between u- and d-quark contributions.
The u- and d-quark contributions to $g_A^{0-}$
are in fact individually small,
which one can find in the columns named as
$g_A^{0-(u)[{\rm lat}]}$ and $g_A^{0-(d)[{\rm lat}]}$
in Table~\ref{fittedval}.
We additionally make some trials 
with lighter u- and d-quark masses at $\kappa$=0.1410.
Since we have less gauge configurations 
and the statistical fluctuation is larger at this kappa,
we fail to find a clear plateau
in the effective mass plots of the two-point correlators
and the extracted mass $E_1^-$ of the negative-parity state 
cannot be reliable.
Leaving aside these failures, we try to extract $g_A^{0-}$.
The result is added in the lower panel in Fig.~\ref{AxialVectorC}
as a faint-colored symbol,
which is consistent with those obtained at other $\kappa$'s.
On the other hand,
the axial charge $g_A^{1-}$ of $N^*(1650)$
is found to be about 0.55,
which has almost no quark-mass dependence.
The striking feature is that
these axial charges, $g_A^{0-}\sim 0$ and $g_A^{1-}\sim 0.55$,
are consistent with the naive nonrelativistic quark model
calculations~\cite{Nacher:1999vg,Glozman:2008vg},
$g_A^{0-}= -\frac19$ and $g_A^{1-}= \frac59$.
Such values are obtained 
if we assume that the wave functions of $N^*(1535)$ and $N^*(1650)$ are 
$|l=1, S=\frac12\rangle$ and $|l=1, S=\frac32\rangle$
neglecting the possible state mixing.
(Here, $l$ denotes the orbital angular momentum
and $S$ the total spin.)

In the chiral doublet model~\cite{DeTar:1988kn,Glozman:2007jt},
the small $g_A^{N^*N^*}$ is realized
when the system is decoupled from the chiral condensate
$\langle \bar \psi \psi \rangle$.
The small $g_A^{0-}$ of $N^*(1535)$ then
does not contradict with the possible and attempting scenario,
the 
{\it chiral restoration scenario in excited hadrons}~\cite{Glozman:2007jt}.
If this scenario is the case, the origin of mass 
of $N^*(1535)$ (or excited nucleons) is essentially
different from that of the positive-parity
ground-state nucleon $N(940)$,
which mainly arises from the spontaneous chiral symmetry breaking.
However, the non-vanishing axial charge of $N^*(1650)$
unfortunately gives rise to doubts about the scenario.

In order to reveal the realistic chiral structure,
studies with much lighter u,d quarks will be indispensable.
A study of the axial charge of Roper,
as well as the inclusion of strange sea quarks could also 
cast light on the low-energy chiral structure of baryons
and the origin of mass.

{\bf Conclusions.}
In conclusion, we have performed 
the first lattice QCD study of the axial charge $g_A^{N^*N^*}$
of $N^*(1535)$ and $N^*(1650)$, with two flavors of
dynamical quarks employing the renormalization-group improved gauge
action at $\beta$=1.95 and the mean-field improved clover quark action
with the hopping parameters, $\kappa$=0.1375, 0.1390 and 0.1400. 
We have found the small axial charge $g_A^{0-}$ of $N^*(1535)$,
whose absolute value seems less than 0.2 
and which is almost independent of quark mass,
whereas the axial charge $g_A^{1-}$ of $N^*(1650)$ is found to be about 0.55.
These values are consistent with the
naive nonrelativistic quark model predictions,
and could not be the favorable evidences for 
the chiral restoration scenario in (low-lying) excited hadrons.
Further investigations on the axial charges
of $N^*(1535)$ or other excited baryons will cast light on the
chiral structure of the low-energy hadron dynamics and 
on where hadronic masses come from.

\acknowledgments
All the numerical calculations were performed
on NEC SX-8R at RCNP and CMC, Osaka University, 
on SX-8 at YITP, Kyoto University,
and on BlueGene at KEK.
The unquenched gauge configurations
employed in our analysis
were all generated by CP-PACS collaboration~\cite{AliKhan:2001tx}.
We thank L.~Glozman, D.~Jido, S.~Sasaki, and H.~Suganuma
for useful comments and discussions.
This work is supported by a Grant-in-Aid for
Scientific research by Monbu-Kagakusho
(No. 17540250), the 21st Century COE
``Center for Diversity and University in Physics'',
Kyoto University and 
 Yukawa International Program for 
Quark-Hadron Sciences (YIPQS).

\end{document}